\documentclass[aps,prl,twocolumn, amsmath,amssymb,longbibliography]{revtex4}

\usepackage{hyperref}
\usepackage{graphicx,graphics,color,epsfig} 
\usepackage{amsmath, amssymb,amsfonts} 
\usepackage{bm,times,xspace,mhchem} 
\usepackage[dvipsnames]{xcolor}

\begin{document}

\title{On the Origin of Acoustic Spin and Elastic Spin:\\Uncovering Hidden Wave Spin of Scalar Fields with Higher-Order Derivative Lagrangian}
\author{Shuo Xin$^{1,2}$}
\author{Jie Ren$^{1}$}
\email{Corresponding Email: Xonics@tongji.edu.cn}
\affiliation{%
$^1$Center for Phononics and Thermal Energy Science, China-EU Joint Lab on Nanophononics, Shanghai Key Laboratory of Special Artificial Microstructure Materials and Technology, School of Physics Science and Engineering, Tongji University, Shanghai 200092, China
}%

\affiliation{
$^2$SLAC National Accelerator Laboratory, Stanford University, Stanford, CA 94309, USA
}

\date{\today}

\begin{abstract}
Scalar field should have no spin angular momentum according to conventional understandings in classical field theory. 
Yet, recent studies demonstrate the undoubted existence of wave spin endowed by acoustic and elastic longitudinal waves, which are of irrotational curl-free nature without vorticity and can be described by scalar fields. 
Here, to solve this seeming discrepancy, we uncover the origin of wave spin in scalar fields beyond traditional formalism by clarifying that the presence of higher order derivatives in scalar field Lagrangians can give rise to non-vanishing spin. 
For scalar fields with only first order derivative, we can make the hidden wave spin emerge, by constructing a latent field that leads to the original field through a time derivative so that is of second order.  
We exemplify the wave spin for elastic and acoustic fields, as well as for dissipative media, following Noether's theorem in higher-order derivative Lagrangian. 
The results would prompt people to build more comprehensive and fundamental understandings of structural wave spin in classical fields.
\end{abstract}

\maketitle





\textit{Introduction.--}
Recently, there has been a rapid sprouting interest on the study of elastic spin \cite{long2018intrinsic,li2019valley,yuan2021observation, sonner2021ultrafast,sasaki2021magnetization,zhao2022elastic, PhysRevLett.129.204303, ren2022elastic, cao2023observation} and acoustic spin\cite{10.1093/nsr/nwz059,bliokh2019spin,bliokh2019transverse,rondon2019acoustic,long2020realization, 10.1093/nsr/nwaa040,wang2021spin,yang2021real,alhaitz2022confined}, after people recognized the existence of spin angular momentum (spin AM) in longitudinal waves \cite{long2018intrinsic, 10.1093/nsr/nwz059}. Both simulations and experiments have demonstrated that the spin AM of longitudinal waves can be observed in various scenarios, such as metamaterials, evanescent waves, and wave interferences \cite{wang2021spin,long2020realization,bliokh2019transverse,toftul2019acoustic,yuan2021observation,calderin2019experimental,ren2022elastic,cao2023observation,10.1093/nsr/nwaa040,10.1093/nsr/nwz059,sonner2021ultrafast,yang2021real,alhaitz2022confined,PhysRevE.105.065208,zhang2020unidirectional, sato2021dynamic, zhao2020phonon, sonner2021ultrafast, rondon2019acoustic}. Its orbital AM counterpart \cite{long2018intrinsic} has also attracted much attention  \cite{deymier2018elastic, nieves2020rayleigh, geilen2020interference,chaplain2022elastic}.

Efforts to understand acoustic and elastic spin have been dedicated to both experimental and theoretical approaches. Numerous studies aim to establish a solid theoretical foundation for the spin of acoustic and elastic waves \cite{long2018intrinsic,10.1093/nsr/nwz059,10.1093/nsr/nwaa040,burns2020acoustic, chaplain2022elastic, bliokh2022field,ren2022elastic,PhysRevLett.129.204303}. However, amidst the abundance of recent research lies the unresolved question of how longitudinal waves, which are scalar in nature, can possess spin AM. Some researchers attribute this phenomenon to the limitations of scalar descriptions \cite{10.1093/nsr/nwz059, du2020echoes, burns2020acoustic}. Indeed, most studies employ a vector field framework when discussing elastic and acoustic waves \cite{burns2020acoustic, bliokh2022field, PhysRevLett.129.204303,ren2022elastic}. Here, however, we demonstrate that a scalar description is not only concise but also capable of explaining the emergence of spin AM.

The assertion that scalar fields do not possess spin AM is a well-established result in classical field theory \cite{soper2008classical,landau1986theory,Landau:1975pou}. However, if we carefully check this statement, we will find it relies on an implicit prerequisite: the Lagrangian of the scalar field is only of the first order, i.e., containing only first-order derivatives. While this is true for most fields that describe fundamental particles in nature, it is not always the case, e.g., for the effective scalar potential $\psi$ of displacement $(\vec x = \vec \nabla \psi)$ in longitudinal elastic and acoustic waves. In these instances, we find that second and third derivatives are involved, which are crucial for the emergence of spin AM. Consequently, our discussions must be based on higher-order Lagrangians.

The study of higher-order Lagrangians has a long history. Notable results, such as Ostrogradsky instability \cite{Ostrogradsky:1850fid}, continue to be discussed in modern studies \cite{motohashi2015third, Langlois:2015cwa}. The Noether currents associated with higher-order Lagrangians have been summarized in literature from the last century \cite{chang_1948, thielheim1967note, durr1974conservation}. The emergence of acoustic spin and elastic spin in scalar longitudinal wave fields, however, was not paid attention there. 

In this Letter, we explain the emergence of spin AM for scalar fields with higher order derivative Lagrangian, and apply the higher order field theory to shearless elastic field, ideal acoustic field and dissipative acoustic field of irrotational curl-free nature. For scalar field that has no spin AM, we can make the hidden wave spin emerge, by constructing a latent field whose time derivative leads to the original field, so that is of higher order. Finally we give remarks on general higher order scalar fields, and conclude in the end.

\textit{Spin Angular Momentum of Scalar Fields.--} Following the Noether's theorem \cite{Noether1918}, the AM density of a field is the conserved current corresponding to rotational symmetry. Usually this AM
separates into two parts: the orbital AM part with extrinsic dependence on the choice of reference origin point, and the intrinsic spin AM part without such dependence. For a typical first order Lagrangian $\mathcal{L} (\phi,\partial_\mu \phi, x_\mu) $, the scalar field $\phi$ has only orbital AM while the vector counterpart can possess additional spin AM. We briefly review the reasoning of this argument. As a clarification of the notation, we will use Latin letters $i,j,k,...$ to denote spatial indices, Greek letters $\mu,\nu,\kappa,\chi,...$ to denote spacetime indices with $\mu = 0$ or $t$ being time index. Note that we do not impose Lorentz covariance since acoustic and elastic fields of our prime interest are non-relativistic in the present context.

For the Lagrangian $\mathcal{L} (\phi,\partial_\mu \phi, x_\mu)$ dependent up to first order derivative of scalar field $\phi$, the action is defined as $S = \int d^4 x \mathcal{L} (\phi,\partial_\mu \phi, x_\mu)$. The evolution of $\phi$ over spacetime is determined by imposing the variation of action $\delta S=0$, the so-called least action principle, which gives us the usual Euler-Lagrangian equation:
\begin{equation}
\label{eq_ELtextbook}
	\frac{\partial \mathcal{L} }{\partial \phi} - \partial_\mu \frac{\partial \mathcal{L}}{\partial (\partial_\mu \phi)}=0. 
\end{equation}
Following the Noether's theorem \cite{Noether1918}, the energy-momentum density is the conserved current corresponding to spacetime translation. Namely, under translation: $x_\mu \rightarrow x_\mu + \delta x_\mu$, we can find the variation of Lagrangian $\delta \mathcal{L} = \partial_\mu ( T_{\mu\nu } ) \delta x_\nu $ with the energy-momentum tensor $T_{\mu\nu} = -\frac{\partial \mathcal{L}}{\partial (\partial_\mu \phi)} \partial_\nu \phi  + \delta_{\mu\nu} \mathcal{L}   $, and the time-space translational symmetry $ \partial_\mu ( T_{\mu\nu } ) = 0 $ gives rise to the conservations of energy $\int_V T_{00}$ and canonical momentum $\int_V T_{0j}$, respectively.

We then consider the Noether current corresponding to spatial rotation along an infinitesimal angle $\vec\theta$: $x_j\rightarrow x_j+\theta_i \epsilon_{jik} x_k$, i.e., the AM. The variation of Lagrangian can be found to be $$\delta \mathcal{L} = \theta_i \partial_\mu \left(  T_{\mu j} \epsilon_{jik} x_k \right).$$ As expected, once the boundary terms vanish at infinity, the rotational symmetry leads to the conserved current that has only the orbital AM part $$J_i = L_i = \epsilon_{ikj} x_k T_{0j}.$$ We can see that the AM is dependent on the choice of reference point $x_\mu$ (so-called orbital) in the absence of additional intrinsic term without $x_k$ dependence. This is the standard conclusion that scalar field has no spin AM, conventional in classical textbooks \cite{soper2008classical, landau1986theory, Landau:1975pou}.

However, we would like to point out that the above derivation is only true for the standard Lagrangian of first order derivative, but invalid for Lagrangian of higher-order derivatives. Suppose we have a second order Lagrangian $\mathcal{L} (\partial_\mu \partial_\nu \psi,\partial_\mu \psi,\psi,x_\mu) $ for the scalar field $\psi$, we will show that the hidden wave spin emerges due to the higher-order derivatives. We will derive the spin and orital AM following Noether's theorem. 
In the presence of second order terms, the Euler-Lagrangian equation from the least action principle is modified as:
\begin{equation}
    \frac{\partial \mathcal{L}}{\partial \psi}- \partial_\nu\frac{\partial \mathcal{L}}{\partial ( \partial_\nu \psi )} + \partial_\mu\partial_\nu  \left(\frac{\partial \mathcal{L}}{\partial (\partial_\mu\partial_\nu \psi)} \right)=0.
\end{equation}
Compared to Eq. \ref{eq_ELtextbook} that is at most second order for canonical first order Lagrangian, now the equation governing dynamics of the field $\psi$ contains third or fourth order derivatives.  
With higher derivatives involved in the dynamics, the quantities we were familiar with from the first order Lagrangian, especially the Noether current, would be modified.

The modified Noether current $T_{\mu\nu }$ of translation invariance is derived in Eq. \ref{eq_Tmunu} in Appendix. Then, if we consider an infinitesimal rotation by an angle of $\vec\theta$, i.e. the variation of $x_j$ is $\delta x_j = \theta_i \epsilon_{jik} x_k$, after some algebra detailed in Appendix, we will find the variation of the Lagrangian takes a different form:
\begin{equation}
\label{eq_deltaL_2order}
    \delta \mathcal{L} = \theta_i  \partial_\mu \left(T_{ \mu j} x_k \epsilon_{jik}  - \frac{\partial \mathcal{L}}{\partial ( \partial_\mu \partial_k \psi )} \partial_j \psi  \epsilon_{jik}  \right)
\end{equation}
where $T_{\mu\nu}$ is the energy-momentum tensor. Compared with classical first order case, an additional second term emerges in the parentheses of Eq. \ref{eq_deltaL_2order}. As such, the conserved current corresponding to AM in rotational symmetry separates into two parts: orbital AM part, the first term with extrinsic dependence on choice of reference points, and the spin AM part, the additional intrinsic term independent on choice of reference points. This is also consistent with $\delta \mathcal{L} = \theta_i \partial_\mu \left(  T_{\mu j} \epsilon_{jik} x_k \right)$ for first order Lagrangian, where the second term vanishes $\frac{\partial \mathcal{L}}{\partial ( \partial_\mu \partial_k \psi )} =0$. In other words, the appearance of higher order derivatives creates intrinsic AM contribution from additional variations of Lagrangian.

From Eq. \ref{eq_deltaL_2order}, we can read off the expression for the spin AM density of scalar fields of second order derivative Lagrangian. Rewriting into a more compact form, we have (an overdot in the expression means time derivative $\partial_t$): 
\begin{equation}
\label{eq_SAM1}
s_i = \frac{\partial \mathcal{L}}{\partial (  \partial_k \dot \psi )} \partial_j \psi  \epsilon_{ijk}, \;\;\;\;\;\; \vec s = \vec \nabla \psi \times \frac{\partial \mathcal{L}}{\partial( \vec \nabla \dot \psi)}.
\end{equation}
One would immediately find this expression analogous to the spin AM density of a vector field $\vec u $, i.e., if we start with first order Lagrangian $\mathcal{L}_{\rm vec}(\vec u, \partial_\nu \vec u)$, we would find spin AM density to be $ \vec s_{\rm vec} = \vec u \times \frac{ \partial \mathcal{L}_{\rm vec} }{ \partial \vec u } $. This observation guarantees consistency when $\vec \nabla \psi=\vec u$, which is the case for longitudinal waves of irrotational curl-free nature in shearless media. 

If we have a third order Lagrangian, following the similar derivation of Eq. \ref{eq_EL3}-\ref{eq_end3} in Appendix,  we would find the spin AM density emerging out of the scalar fields as well, being as:
\begin{equation}
\begin{aligned}
\label{eq_SAM3}
    s_i  =& 2 \frac{\partial \mathcal L}{\partial \psi_{,t \mu k }}  \partial_\mu \partial_j \psi \epsilon_{ijk}   - \partial_\mu \frac{\partial \mathcal L}{\partial \psi_{,\mu t k }}  \partial_j \psi  \epsilon_{ijk} \\
    &\quad +  \frac{\partial \mathcal{L}}{\partial ( \partial_t \partial_k \psi )} \partial_j \psi  \epsilon_{ijk}.
\end{aligned}
\end{equation}
The last term is same as the hidden spin for second order scalar field Lagrangian. Other terms involve third order derivatives and generally have no compact notations to represent how indexes are contracted. Though immediate intuitive interpretation of them are not clear, they are necessary components in e.g. cancelling the dissipative part in Lagrangian, as we would see in the example of viscous acoustic field with dissipation. 




\textit{Wave Spin in Elastic field.--} Consider the elastic wave propagating in a solid object. The local displacement in the solid is a vector field $u_i$. The equation of motion is:
$\rho \ddot u_i = c_{ijkl}\partial_j\partial_l u_k$, 
where $\rho$ is the density and $c_{ijkl}$ is the Christoffel matrix of this material. The Lagrangian density of this system is \cite{landau1986theory, laude2015lagrangian}:
\begin{equation}
\label{eq_lag}
\mathcal{L} = \frac 12 \left( \rho \dot  u_i \dot u_i - c_{ijkl} u_{i,j} u_{k,l} \right), 
\end{equation}
which is a typical Lagrangian for vector field with only first order derivatives. The spin AM corresponding to this Lagrangian is:
\begin{equation}
\label{eq_elasticSAM}
\vec s = \vec u \times \frac{\partial \mathcal{L}}{\partial \dot{\vec u}}= \rho \vec u \times \dot {\vec u}.
\end{equation}
Yet, there is a myth when people separate the displacement field into longitudinal (non-curling, irrotational) and transverse (non-diverging, solenoidal) part, according to Helmholtz decomposition theorem. For the longitudinal (irrotational curl-free) part, the displacement field is given by a scalar potential $\vec u = \vec \nabla \phi $. One may wrongly regard the longitudinal wave as spinless since it is a scalar field and $\vec\nabla\times\vec\nabla \phi=0$. However, as we discussed above, the scalar field with higher order Lagrangian can have nontrivial spin AM. The Lagrangian for the scalar potential is 
\begin{equation}
	\mathcal{L} (\partial_\mu \partial_\nu \phi, x_\mu) = \frac 12 \left( \rho \partial_i \dot \phi \partial_i \dot  \phi  - c_{ijkl} \partial_i\partial_j \phi \partial_k\partial_l \phi \right), 
\end{equation}
which involves second order derivatives of $\phi$. According to Eq. \ref{eq_SAM1}, the spin AM density of the field is
\begin{equation}
    \vec s = \vec \nabla \phi \times \frac{\partial \mathcal{L}}{\partial( \vec \nabla \dot \phi)}= \rho \vec \nabla \phi \times \vec \nabla {\dot\phi}.
\end{equation}
This is exactly the same as Eq. \ref{eq_elasticSAM} after replacing $\vec \nabla \phi$ by $\vec u$. Therefore, it clearly demonstrates that the spin AM of the field results from the local spinning of field polarization (gradient) in time domain, rather than the local circulation (vorticity) of field polarization in space domain \cite{ren2022elastic}. 

\textit{Wave Spin in Ideal Acoustic Field.--} Now let's consider longitudinal waves in shearless fluid, starting from continuity and Navier-Stokes equations \cite{landau2013fluid}:
\begin{align}
    \partial_t \rho + \vec \nabla \cdot (\rho \vec V) = & \,\, 0, \nonumber\\
    \rho \left( \partial_t \vec V + (\vec V \cdot \vec \nabla  ) \vec V \right) =& - \vec \nabla P + \eta \nabla^2 \vec V + \frac{\eta}{3} \vec \nabla (\vec \nabla \cdot \vec V). \nonumber
\end{align}
First we study the non-dissipative ($\eta=0$), adiabatic fluid ($\frac{\delta P}{\delta \rho } = c_0^2 $ with $c_0$ being the speed of sound) in linear regime, i.e. $ \rho = \rho_0 + {p}/{c_0^2} $, $ P = p_0 + p $, $ \vec V = \vec v_0 + \vec v $, with $p, \vec v$ being small perturbations around the background pressure $p_0$ and velocity $\vec v_0$. The background quantities satisfy $\dot \rho_0 + \vec \nabla \cdot ( \rho_0 \vec v_0 ) =0$.

Traditionally, one can introduce a scalar field $\phi$ such that $\vec v = \vec \nabla \phi,\, p = - \rho_0 (\partial_t + \vec v_0 \cdot \vec \nabla  ) \phi$, which automatically satisfy the linearized Navier-Stokes equation. The linearized continuity equation gives the master equation governing the dynamics of scalar field $\phi$ \cite{campos2008waves}:
\begin{equation}
\label{eq_sound}
	\frac{1}{\rho_0} \vec \nabla \cdot (\rho_0 \vec \nabla \phi) - (\partial_t + \vec v_0 \cdot \vec \nabla  ) \frac{1}{c_0^2} (\partial_t + \vec v_0 \cdot \vec \nabla  ) \phi =0.
\end{equation}
This differential equation can be reformulated by least action principle and one arrives at a typical Lagrangian density of linearized scalar field:
\begin{equation}
	\mathcal{L} = \frac {\rho_0} {2}  \left( (\vec \nabla \phi)^2  - c_0^{-2} (\partial_t \phi + \vec v_0 \cdot \vec \nabla \phi)^2 \right),
\end{equation}
which involves only first order derivatives. Following the standard theory after Eq. \ref{eq_ELtextbook}, one will arrive at the conclusion that the spin AM is absent in this field, as in classical textbooks~\cite{soper2008classical, landau1986theory, Landau:1975pou}.

To uncover the hidden spin AM of acoustic field, we need to construct a latent scalar field to make the Lagrangian contain higher-order derivatives, as suggested by above discussions. Here, we give the Lagrangian density of a new scalar field $\psi$ whose time derivative leads to the original scalar field through $\phi = \partial_t \psi$. Then the new Lagrangian density with higher-order derivatives for $\psi$ is written as
\begin{equation}
\begin{aligned}
	\mathcal{L} (\partial_\mu\partial_\nu \psi, x_\mu) = \frac{\rho_0}{2} \left( (\vec \nabla\dot \psi )^2 - c_0^{-2} ( \ddot \psi + \vec{v_0} \cdot \vec \nabla \dot \psi )^2 \right).
\end{aligned}
\end{equation}
The latent scalar field $\psi$ has very clear meaning of potential for the displacement $\vec u= \vec\nabla \psi$, because $\phi = \partial_t \psi$ is the potential for the velocity $\vec v = \dot {\vec u}=\vec\nabla\phi$. 

According to Eq. \ref{eq_SAM1}, the spin AM density of the longitudinal acoustic wave is:
\begin{equation}
	\vec s = \rho_0 \vec u \times \vec v + \frac{ p }{c_0^2} \vec u \times \vec v_0 ,
\end{equation}
where $\vec u= \vec \nabla \psi $ is the displacement field, $\vec v= \partial_t \vec u$ is the velocity field,  $p = -\rho_0 (\partial_t + \vec v_0 \cdot \vec \nabla  ) \dot \psi$ is the pressure field. The physical meaning of this spin AM density would be more evident if we rewrite:
\begin{equation}
    \vec s = \vec u \times \delta ( \rho \vec V )
\end{equation}
with the first order perturbation of local acoustic momentum density $\delta (\rho \vec V) =  \rho_0 \delta \vec V + \delta \rho \vec v_0 =  \rho_0 \vec v + \frac{p}{c_0^2} \vec v_0 $ spinning around the equilibrium position.

\textit{Wave Spin in Dissipative Acoustic Field.--} Next, we turn on dissipation in a static fluid background, i.e. $\eta \neq 0,\, \vec v_0 = 0$, and $p_0,\rho_0$ being constant. The linearized Navier-Stokes equation: $ \rho_0 \frac{\partial \vec v}{\partial t} = - \vec \nabla p + \eta \nabla^2 \vec v + \frac{\eta}{3} \vec \nabla (\vec \nabla \cdot \vec v)$, can be satisfied by introducing a scalar field $\phi$ with $\vec v = \vec \nabla \phi, \quad p = - \rho_0\frac{\partial \phi }{\partial t} + \frac 43 \eta \nabla^2\phi $, and the continuity condition leaves us with the master equation governing the scalar field $\phi$:
\begin{equation}
    \nabla^2\phi - \frac{1}{c_0^2} \frac{\partial^2\phi}{\partial t^2} + \frac 43 \frac{\eta}{\rho_0 c_0^2} \frac{\partial }{\partial t} \nabla^2 \phi  =0 .
\end{equation}
This differential equation can be reformulated by least action principle with Lagrangian density
\begin{equation}
\label{eq_L_diss}
    \mathcal{L} = \frac 12 \rho_0 (\vec \nabla \phi )^2 - \frac{1}{2c_0^2} \rho_0 \dot \phi^2 + \frac 23 \frac{\eta}{ c_0^2} \vec \nabla \phi\cdot \vec \nabla \dot \phi .
\end{equation}
Although this is a Lagrangian of second order derivatives, the spin AM of the velocity potential $\phi$ vanishes, i.e., according to Eq. \ref{eq_SAM1}, one obtains $\vec s = \vec \nabla \phi \times \frac{\partial \mathcal{L}}{\partial( \vec \nabla \dot \phi)} = \frac 23 \frac{\eta }{ c_0^2}\vec \nabla \phi \times \vec \nabla \phi =0  $. This should be also expected physically, because the additional term comes from dissipative viscous force, which shall not give rise to any conserved quantities.

Again we need to consider displacement potential underlying the velocity potential, by digging the latent field $\psi$ whose time derivative leads to the original field $ \phi = \dot \psi $. Lagrangian density of the displacement potential $\psi$ is:
\begin{equation}
    \mathcal{L} = \frac 12 \rho_0 (\vec \nabla \dot \psi )^2 - \frac{1}{2c_0^2} \rho_0 \ddot \psi^2 + \frac 23 \frac{\eta }{ c_0^2} \vec \nabla \dot \psi\cdot \vec \nabla \ddot \psi.
\end{equation}
This Lagrangian has third order derivatives, which can have nontrivial spin AM, as expressed in Eq. \ref{eq_SAM3}. The terms involving third order derivatives for this Lagrangian are
$\frac{\partial \mathcal L}{\partial \psi_{,t i j }} =0$
and
$ \frac{\partial \mathcal L}{\partial \psi_{,t t i }} = \frac 23 \frac{\eta }{ c_0^2} \partial_i \partial_t \psi 
$. Expanding all terms we would also see the contributions from dissipative parts cancel: $ s_i =   \frac{4\eta}{3 c_0^2} \partial_j \partial_t \psi \partial_t \partial_k \psi \epsilon_{ijk}  - \partial_t (\frac{2\eta }{3 c_0^2} \partial_k \partial_t \psi  ) \partial_j \psi  \epsilon_{ijk}  +  (\rho_0 \partial_t \partial_k \psi + \frac{2\eta}{3 c_0^2} \partial_k \partial_t \partial_t \psi ) \partial_j \psi  \epsilon_{ijk}  = \rho_0 \partial_t \partial_k \psi\partial_j \psi  \epsilon_{ijk}$. Therefore, in a compact form, we arrive at:
\begin{equation}
    \vec s =  \rho_0 \vec \nabla \psi \times \vec \nabla \dot \psi  = \rho_0  \vec u \times \vec v .
\end{equation}
This is again consistent with the spin AM of acoustic fields obtained previously, which is not affected by the viscous dissipation.


\textit{Discussions.--} 
For completeness, we also give the general expression of spin AM for scalar field with $N$-th order derivative Lagrangian, following the same approach of Noether's theorem:
\begin{equation}
\label{eq_Nth}
\begin{aligned}
\begin{aligned}
    &s_i = \epsilon_{ijk} \sum^{m+n \leq N-2}_{m, n = 0} (-1)^n (m+1) \times \\
&\underbrace{\psi_{, j \kappa_1 ... \kappa_{m}} }_{(m+1){\rm th\, derivative} } \;
    \underbrace{\partial _ { \chi_1 ... \chi_n } }_{n{\rm th\, derivative}}  
    \frac{\partial \mathcal L}{\partial \psi_{, t k \chi_1 ... \chi_n \kappa_1 ... \kappa_{m}} }.
\end{aligned}
\end{aligned}
\end{equation}
This is similarly obtained in the second line of Eq. 12 in T. S. Chang's 1948 paper on field theories with higher order derivatives \cite{chang_1948} though in slightly different notations. In fact, after we finish the work, we find that the spin AM, as well as other Noether currents of arbitrary higher order Lagrangian was mathematically discussed in notes from last century \cite{chang_1948,thielheim1967note, durr1974conservation}, although without emphasis on the emergence of hidden acoustic spin and elastic spin from scalar fields. 

From Eq.~\ref{eq_Nth}, it is clear that the case of order $N<2$ is trivially $0$ since there is no term to sum over, i.e. classical scalar Lagrangian of first order derivative has no spin AM. While, the cases of  $N=2, 3$ will reduce to the Eqs. \ref{eq_SAM1}, \ref{eq_SAM3}, respectively, which clearly unravel the hidden wave spin of scalar fields from higher-order derivative Lagrangian, explaining the origin of spin AM in irrotational curl-free longitudinal waves. Our representation makes it readily applicable to elastic and acoustic waves represented in modern studies, and the wave spin results are consistent with those from vector field representations \cite{long2018intrinsic,burns2020acoustic,bliokh2022field,PhysRevLett.129.204303,ren2022elastic}. Heuristically speaking, with higher order derivatives, $(N-1)$-th order gradient of $\psi$ would become dynamical coordinate, while the term ${\partial \mathcal L}/{\partial \psi_{, t k \chi_1 ... \chi_n \kappa_1 ... \kappa_{m}} }$ in Eq.~\ref{eq_Nth} plays the role of conjugate momentum, and their antisymmetric contraction with $\epsilon_{ijk}$ would contribute to spin AM density.


\textit{Conclusions.--} As we have seen in this work, the spin AM is simply originated from Lagrangian with higher order derivatives, which invalidates the argument that scalar field has no spin, a result specific to first order Lagrangian. With the help of this insight, we have uncovered the spin AM of longitudinal elastic and acoustic waves, as well as in dissipative media.

This study further reminds us to reconsider typical scalar fields with only first order Lagrangian. We can always construct higher order Lagrangian by introducing a latent field describing the same system after fixing a gauge. Acoustic field is an instance. The Lagrangian for velocity potential $\phi$ is first order, but that for displacement potential $\psi$ with $\phi=\partial_t \psi$ is second order. By introducing this latent field $\psi$, the Lagrangian acquires a gauge symmetry: $\mathcal{L}$ is invariant under transformations of $\psi$ that leave $\phi$ invariant. Here it is simply $\psi \rightarrow \psi + f(\vec x)$ with arbitrary time independent function $f(\vec x)$. The gauge choice has the meaning of $\psi$ at initial time, which is effectively the initial position. By introducing this latent field we uncover the hidden spin angular momentum of acoustic waves. The same logic shall apply to other scalar fields.

This classical picture of Lagrangian mechanics is already enough to see the emergence of spin out of scalar fields. To further understand the spin and other physics of elastic and acoustic waves in this higher order Lagrangian description, one may want to do field quantization to see the behavior of phonon and other elementary excitation in this description. Quantization of higher order Lagrangian can be significantly different from that of first order \cite{dewet_1948,chang_quantum}. This will deserve further exploration in the future.

\begin{acknowledgments}
We acknowledge the support from the National Natural Science Foundation of China (Grant No. 11935010), the Shanghai Science and Technology Committee (Grants No. 23ZR1481200 and 23XD1423800), the National Key R\&D Program of China (Grant No. 2022YFA1404400), and the Opening Project of Shanghai Key Laboratory of Special Artificial Microstructure Materials and Technology.
\end{acknowledgments}

\appendix
\section{Appendix}
\subsection{Derivation of spin AM for second order Lagrangian}

If the Lagrangian $\mathcal L $ has dependence on $x^\mu , \psi , \partial_\mu \psi , \partial_{\mu}\partial_\nu \psi$. 
\begin{equation}
\begin{aligned}
    \delta \mathcal{L} =  \frac{\partial \mathcal L}{\partial (\partial_\mu\partial_\nu \psi )} \delta \partial_\mu\partial_\nu \psi+ \frac{\partial \mathcal L}{\partial (\partial_\mu \psi )} \delta \partial_\mu \psi+ \frac{\partial \mathcal L}{\partial \psi } \delta \psi+ \frac{\partial \mathcal L}{\partial x^\mu} \delta x^\mu
\end{aligned}
\end{equation}

Doing several integration by parts:
\begin{equation}
\begin{aligned}
\delta \mathcal L =& \partial_\mu \left( \frac{\partial \mathcal L}{\partial (\partial_\mu\partial_\nu \psi )} \delta \partial_\nu \psi \right ) - \partial_\nu \left( \partial_\mu \frac{\partial \mathcal L}{\partial (\partial_\mu\partial_\nu \psi )} \delta \psi  \right )\\
& + \partial_\mu \partial_\nu \frac{\partial \mathcal L}{\partial (\partial_\mu\partial_\nu \psi )} \delta \psi \\
& + \partial_\mu \left( \frac{\partial \mathcal{L}}{\partial(\partial _\mu \psi )}  \delta \psi \right) - \partial_\mu \frac{\partial \mathcal{L}}{\partial(\partial _\mu \psi )} \delta \psi \\
& + \frac{\partial \mathcal L}{\partial \psi } \delta \psi + \frac{\partial \mathcal L }{\partial x^\mu } \delta x^\mu 
\end{aligned}
\end{equation}

Using Euler-Lagrangian equation we finally get
\begin{equation}
\begin{aligned}
\label{eq_deltaL}
\delta \mathcal L =& \partial_\mu \left( \frac{\partial \mathcal L}{\partial (\partial_\mu\partial_\nu \psi )} \delta \partial_\nu \psi \right ) - \partial_\nu \left( \partial_\mu \frac{\partial \mathcal L}{\partial (\partial_\mu\partial_\nu \psi )} \delta \psi  \right )\\
& + \partial_\mu \left( \frac{\partial \mathcal{L}}{\partial(\partial _\mu \psi )}  \delta \psi \right) + \frac{\partial \mathcal L }{\partial x^\mu } \delta x^\mu 
\end{aligned}
\end{equation}

For translation
\begin{equation}
    \delta x^\mu  = a^\mu 
\end{equation}
\begin{equation}
    \delta \psi = - \partial_\mu \psi  \delta x^\mu = -\partial_\mu \psi  a^\mu 
\end{equation}
\begin{equation}
    \delta \partial_\mu \psi = \partial_\mu \delta \psi =  -\partial_\mu\partial_\nu  \psi  a^\nu 
\end{equation}
plug into Eq. \ref{eq_deltaL} we get

\begin{equation}
\begin{aligned}
\label{eq_Tmunu}
\delta \mathcal{L} = \partial_\nu \bigg(-\frac{\partial \mathcal{L}}{\partial ( \partial_\nu \partial_\lambda \psi )}  \partial_\lambda\partial^\mu \psi 
+\partial_\lambda \frac{\partial \mathcal{L}}{\partial ( \partial_\nu \partial_\lambda \psi )} \partial^\mu \psi \\
 - \frac{\partial \mathcal L}{\partial (\partial_\nu \psi )} \partial^\mu \psi
+\eta^{\mu\nu} \mathcal{L} \bigg) a_\mu .
\end{aligned}
\end{equation}
The quantity inside the bracket is the energy-momentum tensor $T^{\mu\nu }$.

For spatial rotation
\begin{equation}
    \delta x_i  = \theta_j \epsilon_{ijk} x_k
\end{equation}
\begin{equation}
    \delta \psi = - \partial_i \psi  \delta x_i = -\partial_i \psi  \theta_j \epsilon_{ijk} x_k 
\end{equation}
\begin{equation}
    \delta \partial_i \psi =- \partial_i ( \partial_j \psi  \theta_l \epsilon_{jlk} x_k  ) =   \partial_i  \partial_j \psi  \delta x_j - \partial_j \psi  \theta_l \epsilon_{jli} 
\end{equation}
Plug it into Eq. \ref{eq_deltaL} we get
\begin{equation}
\begin{aligned}
\delta \mathcal{L} = & \partial_\mu \left(-\frac{\partial \mathcal{L}}{\partial ( \partial_\mu \partial_i \psi )} \partial_j \psi \theta_l \epsilon_{jli} \right)  \\
&- \partial_\mu  \bigg(\frac{\partial \mathcal{L}}{\partial ( \partial_\mu \partial_i \psi )}  \partial_j\partial_i \psi \delta x_j
-\partial_\lambda \frac{\partial \mathcal{L}}{\partial ( \partial_\mu \partial_\lambda \psi )} \partial_j \psi\delta x_j  \\
&  + \frac{\partial \mathcal L}{\partial (\partial_\mu \psi )} \partial_j \psi \delta x_j
-\eta^{\mu j} \mathcal{L} \delta x_j\bigg) \\
=& \theta_i  \partial_\mu \left(-\frac{\partial \mathcal{L}}{\partial ( \partial_\mu \partial_k \psi )} \partial_j \psi  \epsilon_{jik} \right) + \theta_i  \partial_\mu \left( \epsilon_{jik} T^{ j\mu} x_k \right)
\end{aligned}
\end{equation}

The first term is independent of the reference point $x_j$. So the conserved charge should be recognized as spin AM $s_i$:
\begin{equation}
	s_i = \frac{\partial \mathcal{L}}{\partial (\partial_t \partial_k \psi )} \partial_j \psi \epsilon_{ijk} 
\end{equation}

This is also the same as Eq. \ref{eq_Nth} with $N=2$. In this case the summation only has one term $m=n=0$, we have $ s_i = - \frac 12 \epsilon_{ijk} (\psi_{,k}\frac{\partial \mathcal{L} }{ \partial \psi_{,tj} } - \psi_{,j}\frac{\partial \mathcal{L} }{ \partial \psi_{,tk} } )  = \epsilon_{ijk} \psi_{,j}\frac{\partial \mathcal{L} }{ \partial \psi_{,tk}} $, same as the equation above.

\subsection{Derivation of spin AM for third order Lagrangian}

When the Lagrangian has dependence up to third order derivatives: $\mathcal L (x^\mu , \psi , \partial_\mu \psi , \partial_{\mu}\partial_\nu \psi, \partial_{\mu}\partial_\nu \partial_\lambda \psi)$, the Euler-Lagrangian equation will be modified by:
\begin{equation}
\begin{aligned}
\label{eq_EL3}
    -\partial_\mu\partial_\nu \partial_\lambda   \left(\frac{\partial \mathcal{L}}{\partial (\partial_\mu\partial_\nu \partial_\lambda  \psi)} \right)
    +\partial_\mu\partial_\nu  \left(\frac{\partial \mathcal{L}}{\partial (\partial_\mu\partial_\nu \psi)} \right) \\
    - \partial_\nu\frac{\partial \mathcal{L}}{\partial ( \partial_\nu \psi )} + \frac{\partial \mathcal{L}}{\partial \psi}=0
\end{aligned}
\end{equation}

The variation
\begin{equation}
\begin{aligned}
    \delta \mathcal{L} =  \frac{\partial \mathcal L}{\partial (\partial_\mu\partial_\nu \partial_\lambda  \psi )} \delta \partial_\mu\partial_\nu \partial _ \lambda  \psi +  \frac{\partial \mathcal L}{\partial (\partial_\mu\partial_\nu \psi )} \delta \partial_\mu\partial_\nu \psi \\
    + \frac{\partial \mathcal L}{\partial (\partial_\mu \psi )} \delta \partial_\mu \psi+ \frac{\partial \mathcal L}{\partial \psi } \delta \psi+ \frac{\partial \mathcal L}{\partial x^\mu} \delta x^\mu
\end{aligned}
\end{equation}

Doing several integrtion by parts  (for simplicity the subscripts $_{,\mu\nu\lambda}$ means $\partial_\mu \partial_\nu \partial_\lambda $)
\begin{equation}
\begin{aligned}
\delta \mathcal L =&
\partial_\mu \left( \frac{\partial \mathcal L}{ \partial \psi_{,\mu\nu\lambda } } \delta \psi_{,\nu\lambda } \right) - \partial_\nu \left(\partial_\mu  \frac{\partial \mathcal L}{ \partial \psi_{,\mu\nu\lambda } }\delta \psi_{,\lambda }   \right)
\\
& +\partial_{\lambda }  \left(\partial_\mu \partial_\nu  \frac{\partial \mathcal L}{ \partial \psi_{,\mu\nu\lambda } }\delta \psi \right) - \partial_\mu\partial_\nu \partial_\lambda  \frac{\partial \mathcal L}{ \partial \psi_{,\mu\nu\lambda } }\delta \psi \\
& +\partial_\mu \left( \frac{\partial \mathcal L}{\partial (\partial_\mu\partial_\nu \psi )} \delta \partial_\mu \psi \right ) - \partial_\mu \left( \partial_\mu \frac{\partial \mathcal L}{\partial (\partial_\mu\partial_\nu \psi )} \delta \psi  \right )\\
& + \partial_\mu \partial_\nu \frac{\partial \mathcal L}{\partial (\partial_\mu\partial_\nu \psi )} \delta \psi \\
& + \partial_\mu \left( \frac{\partial \mathcal{L}}{\partial(\partial _\mu \psi )}  \delta \psi \right) - \partial_\mu \frac{\partial \mathcal{L}}{\partial(\partial _\mu \psi )} \delta \psi \\
& + \frac{\partial \mathcal L}{\partial \psi } \delta \psi + \frac{\partial \mathcal L }{\partial x^\mu } \delta x^\mu 
\end{aligned}
\end{equation}

Using Euler-Lagrangian equation we finally get
\begin{equation}
\begin{aligned}
\label{eq_deltaL3}
\delta \mathcal L =&
\partial_\mu \left( \frac{\partial \mathcal L}{ \partial \psi_{,\mu\nu\lambda } } \delta \psi_{,\nu\lambda } \right) - \partial_\nu \left(\partial_\mu  \frac{\partial \mathcal L}{ \partial \psi_{,\mu\nu\lambda } }\delta \psi_{,\lambda }  \right)
\\
& +\partial_{\lambda }  \left(\partial_\mu \partial_\nu  \frac{\partial \mathcal L}{ \partial \psi_{,\mu\nu\lambda } }\delta \psi \right)
\\
& +\partial_\mu \left( \frac{\partial \mathcal L}{\partial (\partial_\mu\partial_\nu \psi )} \delta \partial_\nu \psi \right ) - \partial_\mu \left( \partial_\mu \frac{\partial \mathcal L}{\partial (\partial_\mu\partial_\nu \psi )} \delta \psi  \right )\\
& + \partial_\nu \left( \frac{\partial \mathcal{L}}{\partial(\partial _\mu \psi )}  \delta \psi \right) + \frac{\partial \mathcal L }{\partial x^\mu } \delta x^\mu 
\end{aligned}
\end{equation}

To see the expression of spin AM, we consider spatial rotation and only keep track of terms that's not orbital part (i.e. not proportional to $\delta x_j$)

For spatial rotation
\begin{equation}
    \delta x_i  = \theta_j \epsilon_{ijk} x_k
\end{equation}
\begin{equation}
    \delta \psi = - \partial_i \psi  \delta x^i = - \partial_i \psi  \theta_j \epsilon_{ijk} x_k 
\end{equation}
\begin{equation}
    \delta \partial_i \psi = \partial_i ( \partial_j \psi  \theta_l \epsilon_{jlk} x_k  ) =  - \partial_i  \partial_j \psi  \delta x_j - \partial_j \psi  \theta_l \epsilon_{jli} 
\end{equation}
\begin{equation}
    \delta \partial_i \partial_j \psi =   - \partial_i \partial_j \partial_k \psi \delta x_k -  \partial_i \partial_k \psi \theta_l \epsilon_{klj} - \partial_j \partial_k \psi \theta_l \epsilon_{kli}
\end{equation}
\begin{equation}
    \delta \partial_t \partial_i \psi =\delta \partial_i \partial_t \psi = - \partial_t \partial_i \partial_k \psi \delta x_k - \partial_t \partial_k \psi \theta_l \epsilon_{kli}
\end{equation}

Plug into Eq. \ref{eq_deltaL3}. The terms that are proportional to $\delta x_j$ don't contribute to spin AM and we don't keep track of them:
\begin{equation}
\begin{aligned}
\label{eq_end3}
    \delta \mathcal{L} =&  \theta_l\partial_\mu \left( 2 \frac{\partial \mathcal L}{\partial \psi_{,\mu i j }}  \partial_i \partial_k \psi \epsilon_{lkj} \right) +  \theta_l \partial_\mu \left( 2 \frac{\partial \mathcal L}{\partial \psi_{,\mu t i }} \partial_t \partial_k \psi \epsilon_{lki} \right)\\
    & -  \theta_l\partial_\mu \left( \partial_\nu \frac{\partial \mathcal L}{\partial \psi_{,\nu \mu i }}  \partial_j \psi  \epsilon_{lji} \right) + \theta_l \partial_\mu \left(\frac{\partial \mathcal{L}}{\partial ( \partial_\mu \partial_i \psi )} \partial_j \psi  \epsilon_{lji} \right)\\
    & + \partial_{\mu} \left( ... \right) \delta x_j
\end{aligned}
\end{equation}

Then we can read off the spin AM as
\begin{equation}
\begin{aligned}
    s_l  =& 2 \frac{\partial \mathcal L}{\partial \psi_{,t i j }} \partial_i \partial_k \psi \epsilon_{lkj}  + 2 \frac{\partial \mathcal L}{\partial \psi_{,t t i }} \partial_t \partial_k \psi \epsilon_{lki} 
    \\
    & - \partial_\nu \frac{\partial \mathcal L}{\partial \psi_{,\nu t i }}  \partial_j \psi  \epsilon_{lji} +  \frac{\partial \mathcal{L}}{\partial ( \partial_t \partial_i \psi )} \partial_j \psi  \epsilon_{lji},
\end{aligned}
\end{equation}
which is exactly  the Eq. \ref{eq_SAM3} by merging the spatial index $i$ and time $t$ into $\mu$. 
This is also the same as the general Eq. \ref{eq_Nth} with $N=3$. In this case the summation has 3 terms for $(m,n)=(0,0),(1,0),(0,1)$. Clearly, we can get the same expression as above:
\begin{equation}
\begin{aligned}
    s_i = & - \frac 12 \epsilon_{ijk} \left(  \psi_{,k} \frac{\partial\mathcal{L}}{\partial \psi_{,tj} } - \psi_{,j} \frac{\partial\mathcal{L}}{\partial \psi_{,tk} } \right)\\
    & - \frac 12 \epsilon_{ijk} \cdot 2 \left(  \psi_{,k\kappa } \frac{\partial\mathcal{L}}{\partial \psi_{,tj\kappa } } - \psi_{,j\kappa} \frac{\partial\mathcal{L}}{\partial \psi_{,tk\kappa } } \right) \\
    & + \frac 12 \epsilon_{ijk} \left(  \psi_{,k} \partial_\chi \frac{\partial\mathcal{L}}{\partial \psi_{,tj\chi } } - \psi_{,j} \partial_\chi \frac{\partial\mathcal{L}}{\partial \psi_{,tk\chi} } \right)\\
    =&  \psi_{,j} \frac{\partial\mathcal{L}}{\partial \psi_{,tk} } \epsilon_{ijk} + 2 \psi_{,j\kappa} \frac{\partial\mathcal{L}}{\partial \psi_{,tk\kappa } } \epsilon_{ijk} -   \psi_{,j} \partial_\chi \frac{\partial\mathcal{L}}{\partial \psi_{,tk\chi} } \epsilon_{ijk}.
\end{aligned}
\end{equation}

\bibliography{references}
\end{document}